\documentclass[prb,preprint,amsmath,amssymb]{revtex4}

\usepackage{graphicx}
\usepackage{bm}
\usepackage{gensymb}
\usepackage{floatrow}
\usepackage{amsmath}
\usepackage{color}

\DeclareGraphicsExtensions{.eps,.jpg,.pdf}

\begin{document}

\title{Nanostructure Investigations of Nonlinear Differential Conductance in NdNiO$_3$ Thin Films}

\author{Will J. Hardy$^1$, Heng Ji$^2$, Evgeny Mikheev$^3$, Susanne Stemmer$^3$, and Douglas Natelson$^{2,4,5}$}
\address{$^1$Applied Physics Program, Rice Quantum Institute, Rice University, Houston, Texas, USA}
\address{$^2$Department of Physics and Astronomy, Rice University, Houston, Texas, USA}
\address{$^3$Materials Department, University of California, Santa Barbara, California, USA}
\address{$^4$Department of Electrical and Computer Engineering, Rice University, Houston, Texas, USA}
\address{$^5$Department of Materials Science and Nanoengineering, Rice University, Houston, Texas, USA}

\date{\today}

\begin{abstract}

Transport measurements on thin films of NdNiO$_3$ reveal a crossover to a regime of pronounced nonlinear conduction below the well-known metal-insulator transition temperature.  The evolution of the transport properties at temperatures well below this transition appears consistent with a gradual formation of a gap in the hole-like Fermi surface of this strongly correlated system. As $T$ is decreased below the nominal transition temperature, transport becomes increasily non-Ohmic, with a model of Landau-Zener breakdown becoming most suited for describing $I(V)$ characteristics as the temperature approaches 2~K. 

\end{abstract}

\maketitle

\section{Introduction}

Strongly correlated materials deviate from the predictions of single-particle band structure calculations that adequately describe many materials. Strong electron-electron and electron-lattice interactions can lead to competition between ground states with very different electronic properties.  Metal-insulator transitions (MITs), often accompanied by structural or magnetic ordering changes, are manifestations of this competition that are of fundamental interest and could potentially prove useful for technological innovation. This category of materials has been the focus of a wide variety of recent investigations \cite{Dagotto2005, Park2013, Radisavljevic2013, Lee2013, Tafti2012} seeking to understand both the physical properties of the correlated state, as well as the complex dynamics that control the electron-electron interactions in solids.

The rare earth perovskite nickelates RNiO$_{3}$, where R is a rare earth ion, are model systems for studying strong correlation behavior. All members of the RNiO$_{3}$ family except for LaNiO$_{3}$ exhibit a metal-insulator transition driven by electronic correlations. \cite{Medarde1997, Son2011} In 1971, Demazeau and collaborators synthesized bulk RNiO$_3$ compounds for the first time and discovered that the electronic and structural properties could be tuned by the choice of rare earth ion. \cite{Demazeau1971, Medarde1997} Interest in such compounds was renewed in the late 1990s when investigations (encouraged by the discovery of high-temperature superconductivity in the cuprate compounds) yielded improved bulk synthesis techniques. \cite{Medarde1997} Although superconductivity was not discovered in RNiO$_3$ compounds, investigators successfully demonstrated control of the MIT by addition of electrons or holes via chemical doping. \cite{Garca-Munoz1995} Due to the difficulty in preparing large, bulk single crystals,\cite{Lorenzo2005} more recent studies have focused primarily on epitaxial thin films, in which parameters such as  material thickness and  strain (via choice of substrate) can be tuned deterministically during growth to modify the material properties. \cite{DeNatale1995, Scherwitzl2010, Xiang2013} 

One such compound, NdNiO$_3$ (NNO), has been the subject of a decades-long debate regarding the mechanism of its metal-insulator transition, specifically whether the low-temperature insulating phase results from formation of a Mott \cite{Stewart2011} or charge-transfer gap,\cite{Torrance1992} or is perhaps due to long-range charge ordering. \cite{Staub2002, Garca-Munoz2009, Hauser2013} Bulk polycrystalline samples exhibit a first-order MIT at $\sim$ 200~K \cite{Lacorre1991, Torrance1992, Granados1993} that separates the orthorhombic, metallic, paramagnetic state at high temperatures from the low-temperature monoclinic, insulating, antiferromagnetic phase. \cite{Garca-Munoz2009, Hauser2013} Once synthesis and doping techniques had been established, investigators sought additional means of altering the MIT. Ionic liquid gating, \cite{Scherwitzl2010, Asanuma2010} as well as modulation doping, \cite{Son2011} have been explored as methods for controlling the transition temperature and conductivity (with one eventual goal of producing a Mott-field-effect transistor), but these have not been shown to completely suppress the MIT in NNO. A recent study \cite{Hauser2013} by Hauser and collaborators employed complementary Hall coefficient and Seebeck measurements to determine that the high-temperature coexistence of a large hole Fermi surface and small electron pocket gives way to a low-temperature pseudogap in the hole Fermi surface, leaving only the electron-like carriers to participate in transport. A forthcoming report from Allen and collaborators details tunneling measurements on thin NNO films that reveal the formation and evolution of a pseudogap at temperatures below $T_{MIT}$. \cite{Allen2014} Further investigation of this low-temperature pseudogap is therefore warranted.

In this work, we employ nano- and micro-structure-based electronic transport to study the nature of the low-temperature insulating state in thin films of NNO. The advantages of this approach are twofold: the film thickness and substrate material can be chosen to tune the lattice strain, and a small lateral distance between deposited metal contacts allows the system to be driven by comparatively high electric fields established by applying modest voltages. Current-voltage ($I(V)$) characteristics encode a wealth of information about the system and make it possible to monitor the temperature-dependent electronic properties.  We observe that a smooth, gradual crossover from Ohmic to non-Ohmic transport accompanies the MIT, and nonlinear transport properties continue to evolve richly at low temperatures, well below $T_{MIT}$. We consider three models to analyze our low temperature experimental $I(V)$ data: Landau-Zener breakdown (LZB), back-to-back Schottky diodes, and space-charge-limited current, and find the best agreement using the LZB model, especially at the lowest temperatures. The data show that contact resistances are not the dominant transport contribution, and that self-heating is unlikely to be responsible for the deviations from Ohmic conduction.  Further studies of nonequilibrium transport in such systems as a function of device orientation and geometry have the potential to shed light on gap formation in correlated materials.

\section{Methods}

Epitaxial NNO thin films were grown by RF magnetron sputtering on LaAlO$_3$ (LAO) substrates, which present a small lattice mismatch of approximately $-0.5\%$ for NNO, resulting in compressive strain. \cite{Catalan2000, Scherwitzl2010} The samples were characterized by x-ray diffraction (XRD) and high-angle annular dark field scanning transmission electron microscopy (HAADF-STEM) to verify the film thickness and crystallinity, similar to a previous work. \cite{Hauser2013} Three film thicknesses were studied: 3.85~nm, 10~nm, and 16~nm. Planar multi-contact devices were defined by electron-beam lithography (EBL), followed by electron-beam evaporation of metal contacts (2~nm Ni and 50~nm Au) and liftoff. A second step of EBL defined a protective mask for reactive ion etching with BCl$_3$, or wet etching with hydrochloric acid. For the samples prepared by wet etching, the contacts used for electrical measurements were deposited after the etching step in order to avoid acid damage to the Ni adhesion layer. The resulting device geometry consisted of a narrow ($\sim$ $5-10$ $\mu$m) strip of film, isolated from its surroundings by the insulating substrate, and several pairs of interdigitated contacts with separation lengths between $\sim$ 100~nm and 6 $\mu$m. A scanning electron microscopy (SEM) image of such a device is shown in the inset of Fig.~\ref{resistivity}.

These devices were measured in a Quantum Design Physical Property Measurement System (PPMS) using low-frequency lock-in techniques as well as dc characterization. Measurement of the temperature-dependent resistivity allowed estimation of $T_{MIT}$. Simultaneous current-voltage ($I(V)$) and differential conductance ($dI/dV$ vs. $V$) measurements were performed by superimposing a small ac excitation on a variable dc bias using a summing amplifier.

\section{Results}

\begin{figure}
\includegraphics[width=8.5cm,clip]{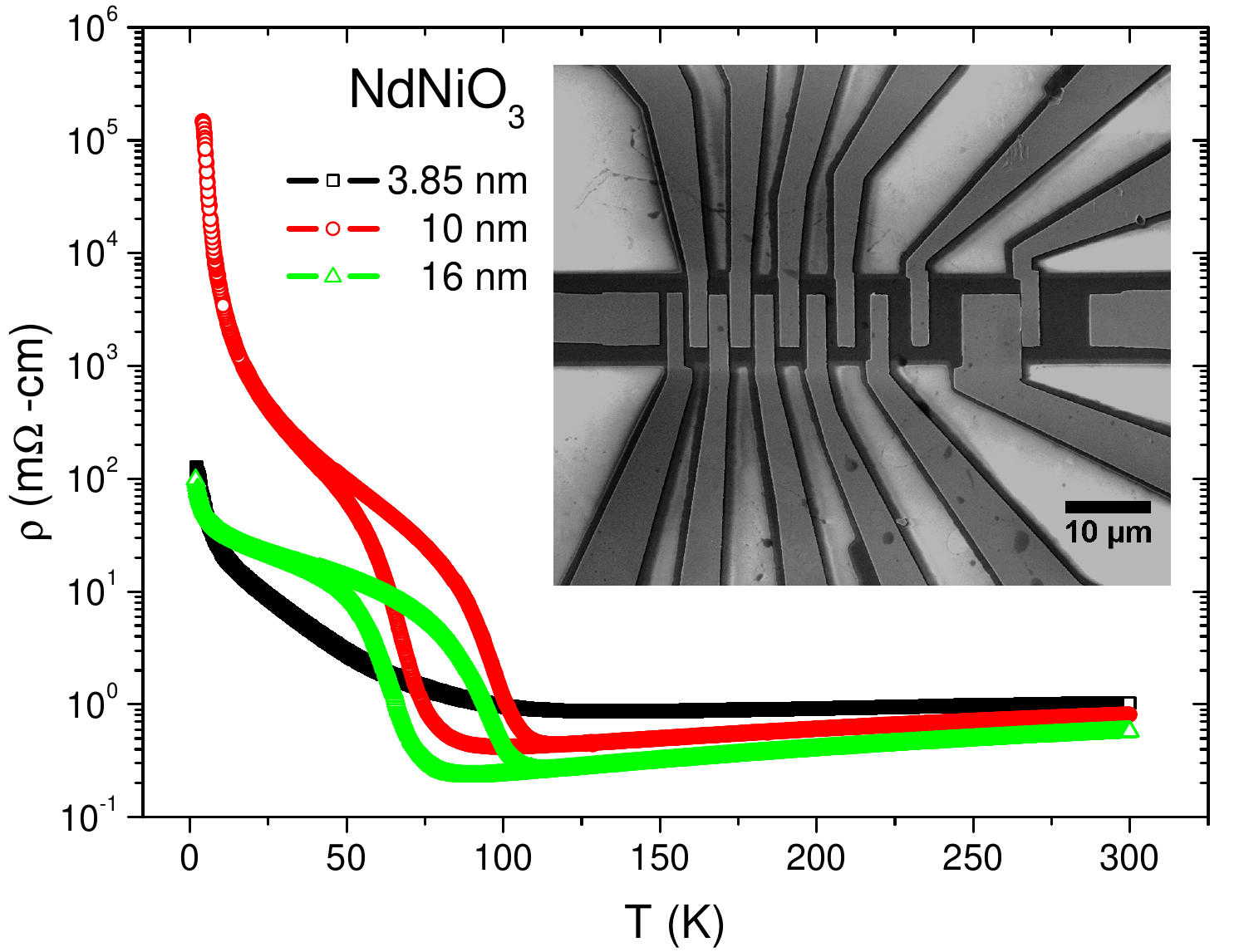}
\caption{\label{resistivity} Temperature-dependent resistivity of NNO films of three different thicknesses. Each sample undergoes a metal-insulator transition upon cooling. When the samples are subsequently warmed, a clear hysteresis is observed in the 10~nm and 16~nm films' data, while the hysteresis in the 3.85~nm data is comparatively small and not visible at this scale. A second upturn is present in each curve below $T_{MIT}$, near 5~K. Inset: SEM image of a representative device with interdigitated contacts. The dark gray horizontal bar is the NNO film; the medium gray areas are the Ni/Au contacts; the light gray area is the LAO substrate where the NNO film was etched away. The scale bar represents 10~$\mu$m.}

\end{figure}

Temperature-dependent resistivity data for three film thicknesses, collected using four-probe measurements, are shown in Fig.~\ref{resistivity}. Between room temperature and $\sim$ 100~K, the resistivity decreases with increasing thickness for the three films (from $\sim$1 m$\Omega$-cm to ~0.5 m$\Omega$-cm at 300~K), and a positive linear slope as a function of temperature in this region is consistent with metallic conduction. An upturn in each curve marks $T_{MIT}$, below which the slope $\tfrac {d\rho} {dT}$ is negative down to low temperatures, indicating insulating behavior. Taking $T_{MIT}$ to be the temperature at which the resistivity reaches its minimum value upon cooling from room temperature, the transition temperature values for each thickness are approximately 138~K (3.85~nm film), 100~K (10~nm film), and 90~K (16~nm film). The reduced value of $T_{MIT}$ compared to the bulk value and the observed smoothness of the MIT are characteristic of thin film samples\cite{DeNatale1995, Scherwitzl2010} and are likely due to the strain imposed by the substrate.   Thin, coherently strained rare earth nickelates are known to have strongly modified Ni-O-Ni bond angles, as a result of interfacial connectivity and lattice mismatch strain, which affects parameters such as band width, resistivity, and correlation physics.\cite{Rondinelli2012,May2010,Hwang2012,Ouellette2010}

At intermediate temperatures, a clear hysteresis between the warming and cooling curves is visible for the 10~nm and 16~nm samples, while the hysteresis is comparatively small for the 3.85~nm sample, probably due to its greater sensitivity to strain effects.\cite{DeNatale1995} At $T$ = 2~K, the 3.85~nm and~16~nm films have similar resistivity values on the order of 100 m$\Omega$-cm; it is unclear why the 10~nm film's resistivity at this temperature is so much larger.  Different deposition conditions were employed for the growth of the 10~nm film (lower sputtering pressure).  The MIT is known to be highly sensitive to both deposition and post-deposition processing procedures,\cite{Hauser2014} likely the reason for the differences between the responses of the 10~nm and 16~nm films.

\begin{figure}
\includegraphics[width=8.5cm,clip]{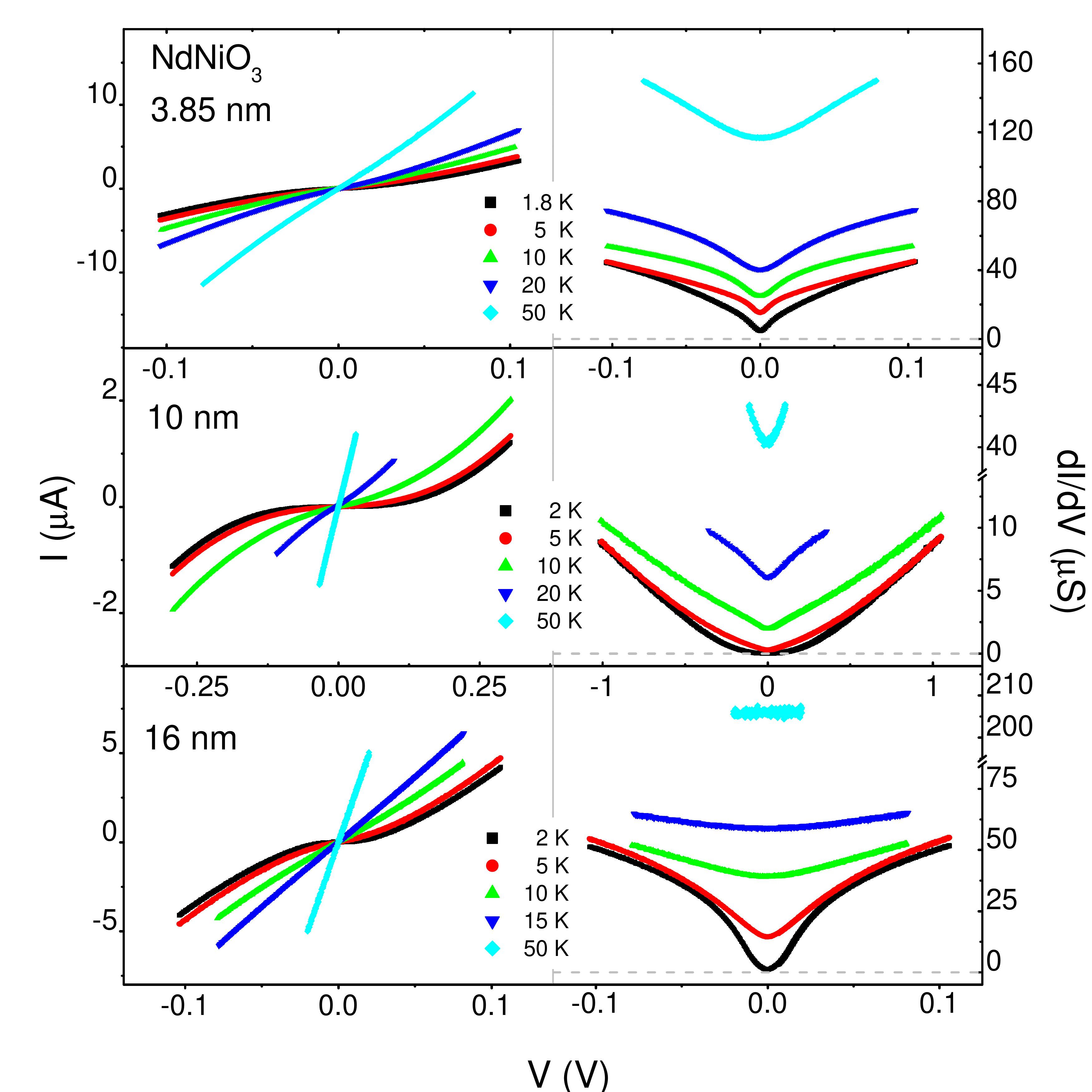}
\caption{\label{IV_and_dIdV} $I(V)$ (left column) and $dI/dV$ vs. $V$ (right column) isotherms for 3.85~nm, 10~nm, and 16~nm film thicknesses at selected temperatures. Below $T_{MIT}$, as the temperature is reduced, $I(V)$ becomes increasingly nonlinear, and $dI/dV$ vs. $V$ develops a dip around zero bias. The specific shape of this dip evolves with temperature and depends upon the sample thickness, likely due to the different strain configurations. The observed suppression of the conductance near zero bias may be due to the formation of a pseudogap that grows with decreasing temperature. }
\end{figure}

To study the transport properties in the insulating state, $I(V)$ and $dI/dV$ vs. $V$ curves were collected at selected temperatures, with particular focus on temperatures below $\sim$ 20~K. $I(V)$ curves for three film thicknesses measured at selected temperatures are shown in the left column of Fig.~\ref{IV_and_dIdV}, along with their corresponding $dI/dV$ vs. $V$ curves in the right column. Above $T_{MIT}$, the $I(V)$ curves are Ohmic, as expected for a system with metallic resistivity properties, but become increasingly nonlinear as the temperature is reduced below $T_{MIT}$. When the temperature is above $\sim$ 50~K, the practically accessible range of bias voltages is typically very small if the current is to be limited to several $\mu$A to avoid detectable Joule heating. High-field data at these temperatures are dominated by self-heating and are therefore not shown.

At sufficiently low temperatures that the temperature hysteresis is negligible, the $I(V)$ curves show rich evolution of their nonlinearity as the temperature is varied, even though one might naively expect that the material would be nearly fully gapped. Previous work has suggested the coexistence of metallic and insulating domains below $T_{MIT}$, with conduction made possible by percolation through metallic domains.\cite{Catalan2000, Hauser2013} A second upturn in the temperature-dependent resistivity curve below $T_{MIT}$ testifies to a possible loss of conduction channels. There is, however, no evidence of conduction taking place through discrete domains, as was previously reported in, \textit{e.g.}, vanadium dioxide films\cite{Sharoni2008} or manganite nanowires,\cite{Zhai2006} other strongly correlated systems with pronounced metal-insulator transitions.

The differential conductance $dI/dV$ allows closer scrutiny of this nonlinear behavior. The $dI/dV$ for each film thickness is suppressed about $V = 0$ when $T < T_{MIT}$, and the suppression becomes stronger with decreasing temperature. The 3.85~nm film has such a dip in the conductance which becomes noticeably sharper as the temperature is reduced from 50~K to 1.8~K. Even at this low temperature, $dI/dV$ is not flat near zero bias, but rather has a relatively sharp minimum at $V$ = 0. The 16~nm film's dip near $V$ = 0 deepens smoothly with decreasing $T$, starting at 10~K down to 2~K. Its $dI/dV$ curve is also not flat at 2~K near zero bias, but instead has a rounded dip. The 10~nm film, which is significantly more insulating at low temperatures than the other two films, exhibits markedly different $dI/dV$ behavior. As the sample is cooled, $dI/dV$ has a smoothly rounded dip at 20~K; at 10~K, there is instead a sharp cusp at zero bias; at 5~K, a flatter region near $V$ = 0 appears; and at 2K, the curve has two well-defined regions at low bias, an apparent gap flat below $\sim$230~mV, and a roughly linear voltage dependence at high bias (above $\sim$1~V), connected by smooth curves. The zero bias suppression of the conductance is suggestive of the formation of a pseudogap as the temperature is reduced.

Self-heating is always a concern in low temperature measurements of nanostructures at significant finite biases.  Strong self-heating in the presence of a strongly temperature-dependent resistivity, as is common in metal-insulator transitions, can lead to dramatic effects, including hysteresis in the apparent dc $I(V)$ response.\cite{Gurevich1987, Altshuler2009, Fursina2009}  We observe no such bistability here. 

In the presence of strong local heating, one would expect the effective average sample temperature within the device channel, $T_{\mathrm{eff}}$, to exceed significantly the nominal substrate temperature, $T_{0}$.  $T_{\mathrm{eff}}$ at a given bias should ``float'' to the value where thermal transport out of the device balances locally dissipated power.  As $T_{0}$ is reduced, the thermal path is expected to become progressively worse due to thermal boundary resistances and the decreasing thermal conductivity of the sample, substrate, and wiring.  All other things being equal, a fixed amount of dissipated power would be expected to lead to at least the same $T_{\mathrm{eff}}$.  Thus, one would expect $I(V)$ curves to saturate or cross at high bias as $T_{0}$ is reduced if the nonlinearity were dominated by self-heating effects, in contrast to the experiments.  Moreover, from the temperature dependence of the zero-bias conductance, we can estimate what change in $T_{\mathrm{eff}}$ would be required at high bias to account for the observed nonlinearity.  We find changes of tens of Kelvin, inconsistent with the actual temperature evolution of the $I(V)$ curves and the lack of saturation of $I$ (at high bias) vs. temperature at low temperatures.  Thus, we conclude that local self-heating is unlikely to play a dominant role in the observed nonlinear effects.

\begin{figure}
\includegraphics[width=8.5cm,clip]{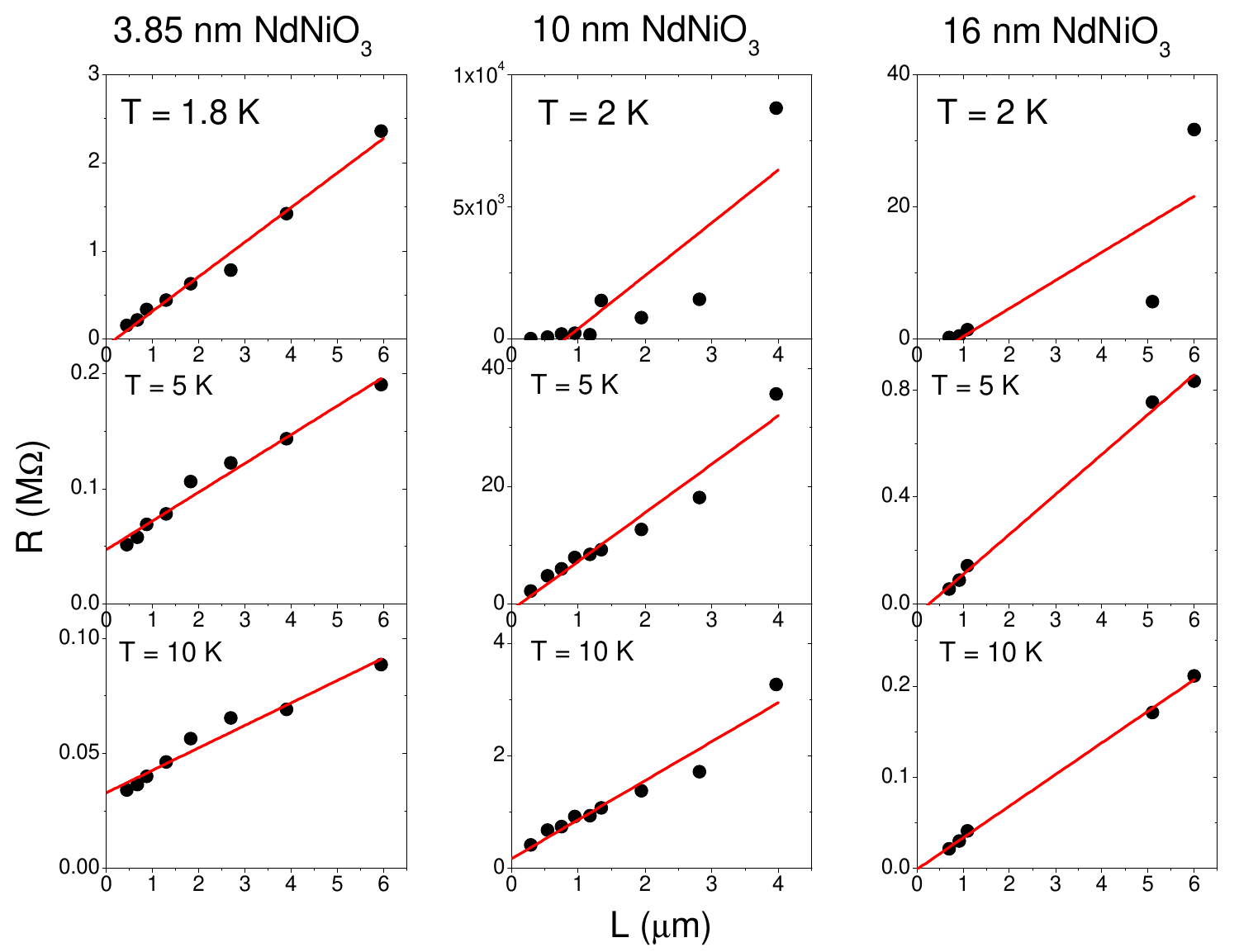}
\caption{\label{R_vs_L} Zero-bias resistance as a function of device length for each film thickness at selected temperatures. The red lines are linear fits to the experimental data (full circles). A superlinear trend is visible at the lowest temperatures measured for each thickness; the data become more linear at higher temperatures.}
\end{figure}

It is also important to determine the contribution of (possibly non-Ohmic) contact resistance to the observed zero-bias conductance suppression and nonlinearity.  There are two approaches available: four-terminal measurements, which become difficult for very resistive samples; and the transmission line method.\cite{Hamadani2004, Reeves1982} The present multi-contact sample design is readily compatible with the latter approach, permitting study of the films' electrical properties as a function of inter-electrode distance (the device length, $L$). Fig.~\ref{R_vs_L} shows plots of zero-bias resistance (extracted from the $I(V)$ curves) as a function of device length for each film thickness at $T$ = 2, 5, and 10~K. Since these data were collected using a two-terminal method, there is some contribution from contact resistance, the total of which corresponds to the $L \rightarrow 0$ intercept. Note that this intercept is small compared to the two-terminal zero-bias resistance at the lowest temperatures.

With perfectly homogeneous device material, every contact interface identical, and Ohmic contact in the bulk and at the contacts, the expectation would be for a linear trend and a positive $y$-intercept in plots of the zero-bias resistance versus measured electrode spacing.  However, the linear fit shown in Fig.~\ref{R_vs_L} clearly deviates from some of the low-$T$ data, and has an unphysical negative $y$-intercept in some cases. The deviations must therefore arise from a combination of non-identical contact resistances at distinct contacts, inhomogeneities in the etched film properties on scales of one micron or larger following the etching process, and potentially deviations in Ohmic length dependence of resistance ($e.g.$, superlinear length dependence of resistance due to localization).  

The negative $y$-intercept observed in some of the linear fits results from scatter in zero-bias resistance values for the longest devices, indicating that the uncertainty in the contact resistance (inferred from the $y$-intercept) is larger than the apparent contact resistance itself.  This situation is not unprecedented, arising frequently in organic semiconductor device measurements,\cite{Xu2010} due to inhomogeneity of the organic channel. Based on our experiences in device fabrication with this material, inhomogeneities in both etched film properties and contact resistances are possible.   In the devices where these concerns are greatest (the 10~nm and 16~nm thicknesses, showing the largest deviations) we note that the data at higher temperatures (5~K, 10~K) are nearly linear with near-zero intercept, consistent with small contact resistances relative to the bulk contribution.  At 2~K, in both cases the longest devices show large deviations from linear length dependence.  We further note that in the data for the 3.85 nm film devices, the contact resistance contribution clearly becomes $less$ important relative to the bulk as temperature is decreased, as seen by the decreasing intercept as $T$ is reduced. In light of this interpretation, it is reasonable to conclude that the two-terminal $I(V)$ and $dI/dV$ vs. $V$ curves are indeed primarily measurements of the bulk NNO channel itself and not of the contacts.

\section{Discussion}

The complex structure of the $I(V)$ curve nonlinearity clearly contains much embedded information about the electronic structure and transport properties of the NNO material as the insulating state becomes robust at low temperatures.  Unlike tunneling spectroscopy, a finite-bias technique known under certain circumstances to measure the local density of states, there is not a \textit{general} theoretical description of finite-bias non-Ohmic transport to allow extraction of the underlying physics.  However, there are some specific transport mechanisms to consider as a starting point for analysis.   Three $I(V)$ fitting models were considered, each linked to a distinct transport  mechanism: (1) a model of Landau-Zener breakdown of the insulating state; (2) a back-to-back Schottky diode model; and (3) a space-charge-limited current model. Our results are most consistent with (1), while attempts to use (2) and (3) did not produce reasonable fits.  Even model (1) is only appropriate at low bias and low temperature, becoming less accurate as $V$ or $T$ increases. 

The Landau-Zener breakdown model uses the following expression (after the result of Oka and Aoki \cite{Oka2005}) to approximate $I$ vs. $V$:

\begin{equation}
\\
I = -A V ln(1 - e^{\tfrac{-V_0} {V}})
\\
\end{equation}

The physics behind this model is the field-driven breakdown of a correlation-based gapped state.  A charge carrier traversing a unit cell acquires enough energy from the dc electric field to overcome the correlation gap and form a mobile excitation of the correlated system.  This is in analogy with Landau-Zener breakdown of the band gap in an ordinary band insulator.  The expectation is that this model should be most appropriate when the system is fully gapped, with ordinary Ohmic transport suppressed.

\begin{figure}[h!]
\includegraphics[width=8.5cm,clip]{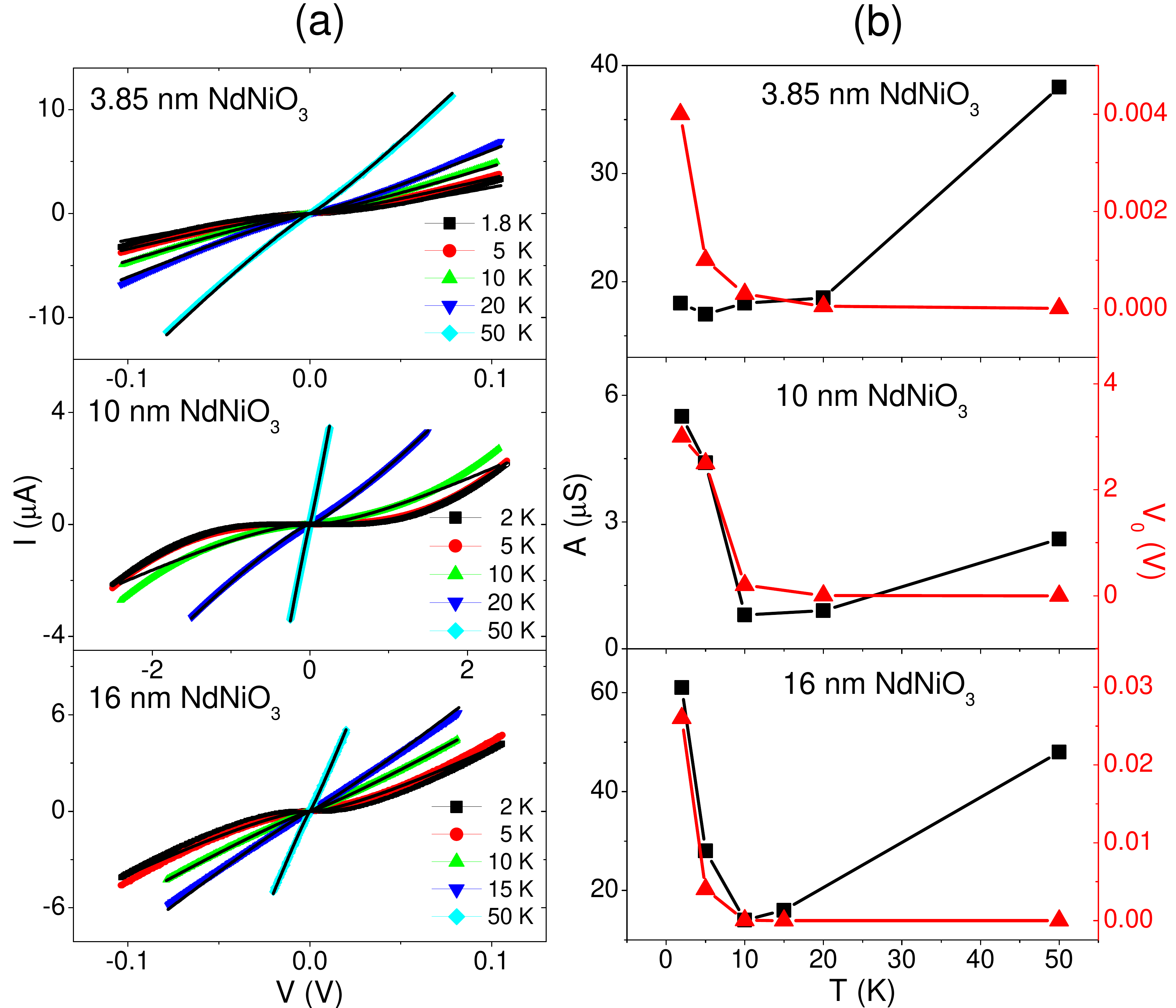}
\caption{\label{IV_fits} (a) Landau-Zener breakdown model fits (solid lines) of experimental $I(V)$ data (full symbols) for each film thickness at selected temperatures. (b) Fit parameters $A$ (black squares) and $V_0$ (red triangles) as a function of temperature for each film thickness.   }
\end{figure}

\begin{figure}[h!]
\includegraphics[width=8.5cm,clip]{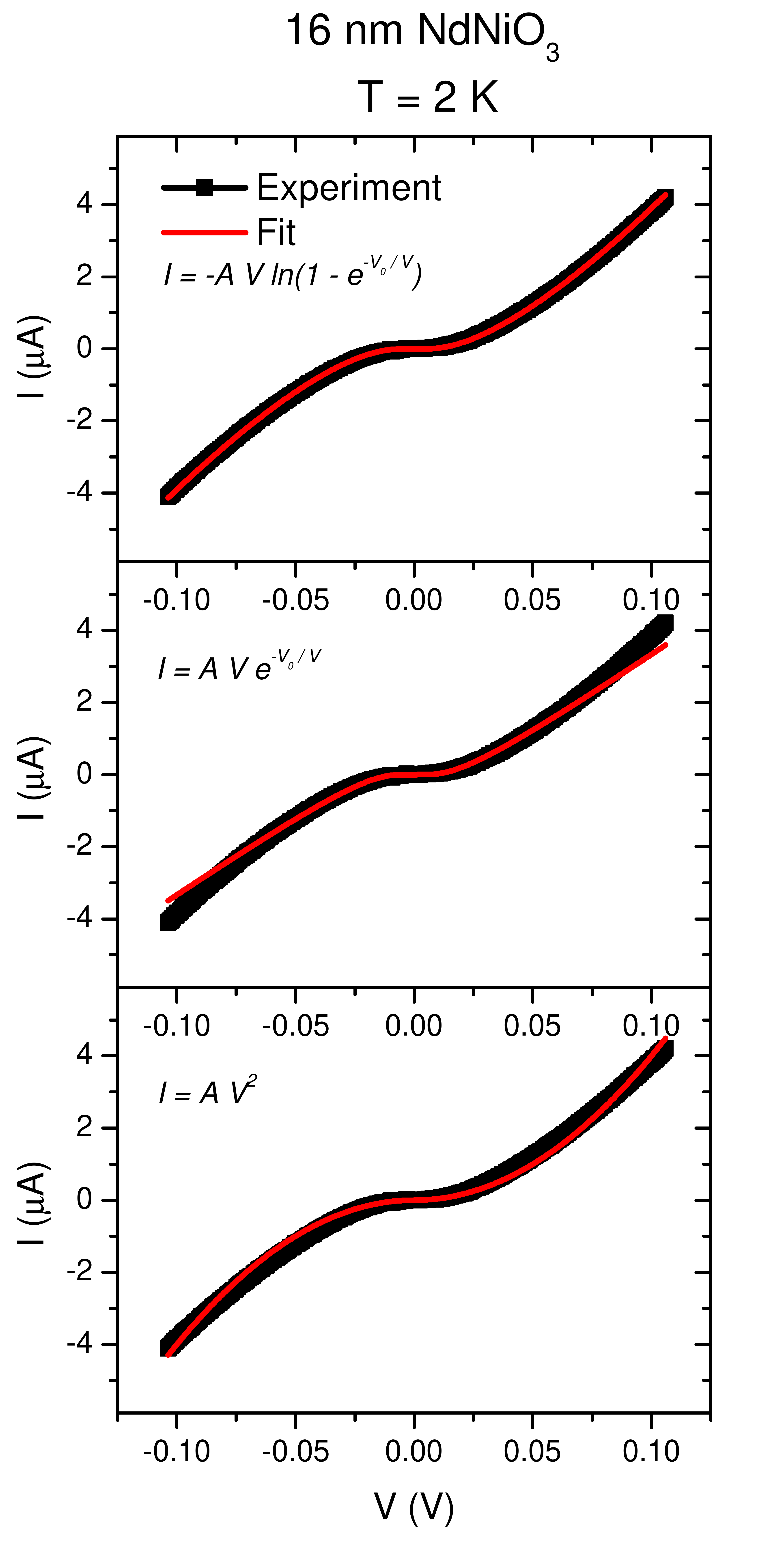}
\caption{\label{comparison_of_fits} Examples of $I(V)$ curve fits to 16~nm film data using three trial models: The LZB model of Oka and Aoki,\cite{Oka2005} an alternative LZB model of Eckstein \textit{et al}.,\cite{Eckstein2010} and a space-charge-limited current model.\cite{Grinberg1989} The first LZB fit is qualitatively the best; the second LZB fit is nearly as good at low bias but noticeably fails at high bias, and the space-charge-limited current fit systematically fails at both low and high bias.   }
\end{figure}

The scaling factor $A$ sets the overall curve slope, while $V_0$ represents a threshold voltage (or electric field), above which the current increases rapidly. Fig.~\ref{IV_fits}a shows fits of the experimental $I(V)$ curves using this model for each film thickness at selected temperatures. Fitting parameters $A$ and $V_0$, which are plotted against temperature in Fig.~\ref{IV_fits}b, were chosen for best agreement between the fit and the experimental data at low bias. The model typically provides the best fit at low bias and low temperatures, becoming less satisfactory at higher bias and temperatures. The overall quality of the fits was approximately the same for all film thicknesses. The model's assumption of a threshold voltage (or field), below which there is very limited conduction, is only consistent with the experimental $I(V)$ curves at the lowest temperatures (near 2~K). At higher temperatures (5~K and above), the slope of the $I(V)$ curve is non-zero and approximately linear close to zero applied bias. This suggests that the Fermi surface gapping is incomplete at temperatures above approximately 2~K, and the kink found in the resistivity curves near $\sim$5~K may mark a crossover to the LZB regime. It is important to note that while other LZB models for $I(V)$ exist, model (1) provides the best agreement with this experiment.  (As an example of such an alternative, a model of LZB presented by Eckstein \textit{et al}.,\cite{Eckstein2010} is found to be less adequate than (1), as shown briefly in Fig.~\ref{comparison_of_fits}).

An alternative model worth considering when dealing with contacts between metals and gapped material is that of Chiquito \textit{et al}., which describes back-to-back Schottky diodes and has been reported to explain the $I(V)$ behavior of one-dimensional semiconducting SnO$_2$ nanobelts contacted by various metals.\cite{Chiquito2012} In this case, the nonlinear effects observed are primarily due to the work function of the metal used to contact the semiconductor, which is a situation well described as a Schottky interface, at the two ends of a narrow strip of an otherwise ballistic conductor.  The model can adequately account for a range of work functions for commonly used contact metals, as well as describe asymmetric behavior that results from microscopic differences between two nominally identical contacts on the same device, for samples measured at and above room temperature. When we attempted to adapt this result to model our $I(V)$ curves obtained at temperatures below 100~K, we found that a kink in the fit near the origin (which is small when the temperature parameter is of order 300~K) becomes the dominant feature over the relevant range of bias voltage (~$<$~1~V). We conclude that this model is not valid for the present situation.  Given that the transmission line analysis demonstrates that contacts do not dominate the $I(V)$ characteristics at low temperatures, it is not surprising that a contact-based model does not describe the data well.

Finally, another model that is often physically relevant when considering charge transport through a nominally insulating material is that of  space-charge-limited current.  In this situation, transport in the bulk is poor enough that the injected charge is distributed nonuniformly and self-consistently modifies the local electric field.  For the thin film case, Grinberg \textit{et al}. calculate that the current density $J$ should be proportional to $V^2$. \cite{Grinberg1989}  When applied to the present case, this model systematically fails to match the experimental data, with the fit falling below at low bias and above at high bias (see Fig.~\ref{comparison_of_fits}). 

One further consideration regarding the nonlinear conduction is whether the relevant parameter is the voltage bias or the electric field.  In the derivation of the LZB model, the physical basis for the breakdown is local electric field (in the form of the voltage drop across a typical lattice length scale). In the limit that transport is a minor perturbation on the potential distribution ($i.e.$, we are not in the space-charge-limited regime, nor is it the case that enough current is flowing that a large fraction of the potential is dropped across some sort of contact resistance of questionable geometric size), one would expect indications that the electric field $E$ is the relevant driving parameter.

\begin{figure}
\includegraphics[width=8.5cm,clip]{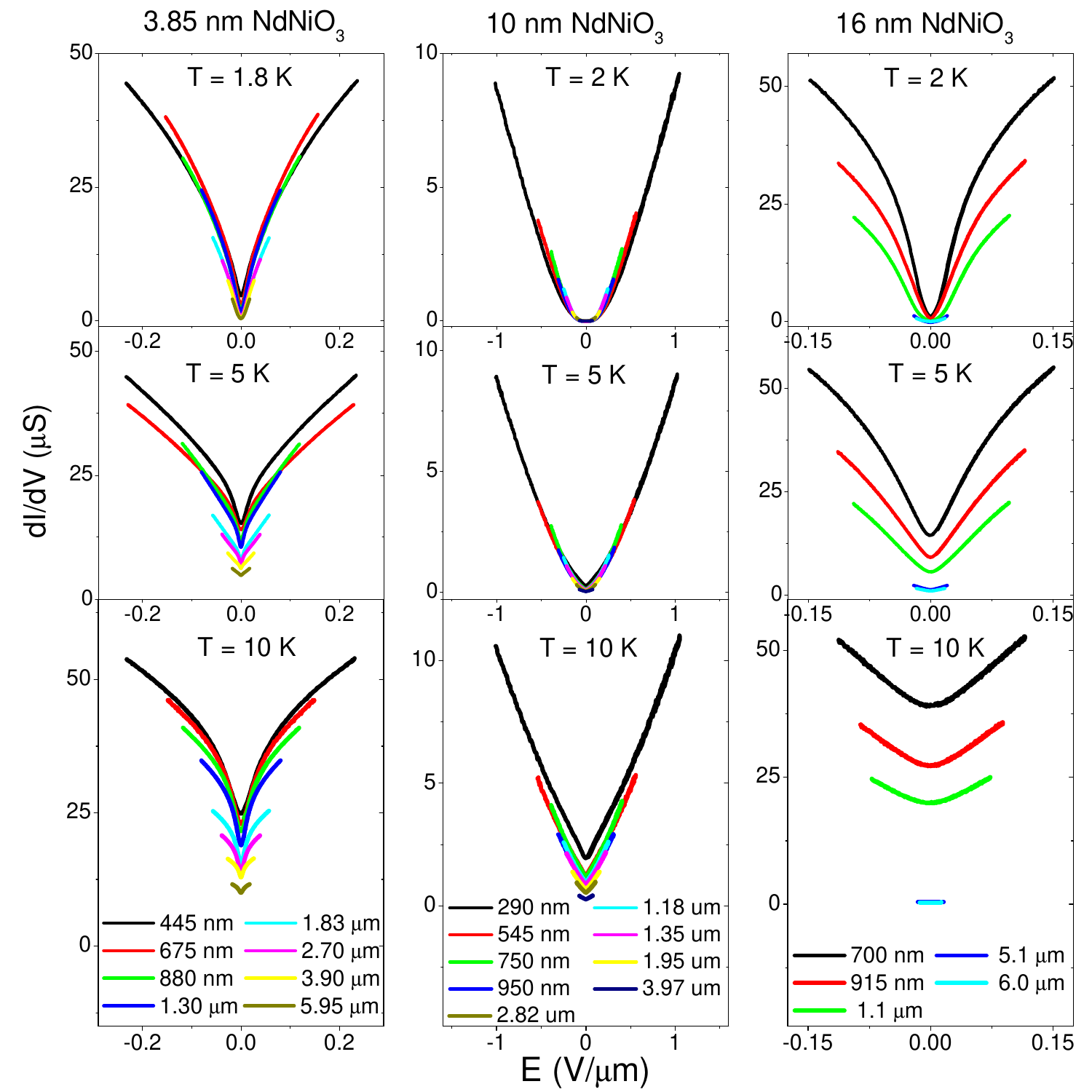}
\caption{\label{dIdV_E} Plots of $dI/dV$ vs. $E$ for each sample thickness at $T$ = 1.8 or 2~K, 5~K, and 10~K. When plotted against $E$, the curves for the 3.85~nm sample at $T$ = 1.8~K and for the 10~nm sample at $T$ = 2~K and 5~K overlap well. This overlap disappears at higher temperatures, and does not appear for the 16~nm sample at any measured temperature.  }
\end{figure}

To test for this, $dI/dV$ was plotted as a function of dc electric field $E$, defined as the dc bias divided by the measured inter-electrode distance.  Plots of $dI/dV$ vs. $E$, shown in Fig.~\ref{dIdV_E}, demonstrate that the data from various gap sizes collapse onto a common curve shape for the 3.85~nm sample at $T$ = 1.8~K and for the 10~nm sample at $T$ = 2~K and 5~K, but not for higher temperatures, or for the 16~nm sample at any measured temperature. The 16~nm sample, which likely has comparatively lower strain overall, may require even lower temperatures for the pseudogap to fully develop.

Plotting the data this way tacitly assumes that the contact resistance is negligible, and that the measured $dI/dV$ is limited by the bulk of the respective channels.  This is equivalent to assuming that of the total applied voltage, the vast majority drops across the channel rather than over some short distance scale at the metal/film contact.  (Note, too, that while plotting conductance in this way, it is not easy to resolve the anomalously high zero bias resistance cases in the 10~nm and 16~nm samples at 2~K, as these have zero-bias conductances that are very small on these scales.)

The observed overlap in $dI/dV$ vs. $E$ supports the idea that the electric field is the parameter driving conduction in the nearly-fully-gapped, low-temperature state in the presence of sufficiently large strain. Although this is not a conclusive proof, the fact that the data do collapse reasonably well this way indicates that the assumptions (contact resistance is not a major contributor to nonlinearity; bulk transport is sufficiently weak at the lowest temperatures that it makes sense to think of the bulk as ``insulating'' and thus capable of exhibiting breakdown-like physics) are self-consistent. 

\section{Conclusions}

We have explored the electronic properties of NdNiO$_3$ thin films via nano- and micro-scale transport measurements.  These experiments show a nonlinear transport regime below the nominal  metal-insulator transition temperature.  Scaling of the two-terminal transport as a function of device length shows that this nonlinearity is not dominated by contact effects, particularly at the lowest temperatures examined.  The temperature and bias evolution also appear to be incompatible with local self-heating playing a dominant role.  The evolution of the $I(V)$ and $dI/dV$ vs. $V$ curves suggests that gapping of the Fermi surface upon cooling is not complete until the temperature is well below $T_{MIT}$.  In the limit that bulk Ohmic transport is suppressed, the $I(V)$ curves are reasonably described by a model of Landau-Zener breakdown in a correlated system.  Consistent with this model, the data in the two most insulating samples at the lowest temperatures suggest that the parameter driving the nonlinear conduction is the electric field rather than the bias voltage.

The rich evolution of nonlinear transport in NNO films demonstrates that much information is encoded in such data.  No general theoretical treatment exists for the analysis of such experiments in the presence of possible contact effects, various bulk transport mechanisms, and extrinsic issues such as local heating.  In particular, there is no model for the crossover from Ohmic conduction to LZB as transport is gapped out via, \textit{e.g.}, decreasing temperature.  However, we believe that systematic studies can lead to insights, provided the parameter space of nonlinear processes may be narrowed (for example, demonstrating that contact effects are not dominant).  Additional experiments that could provide further insight include:  pulsed measurements to quantify any possible heating contribution to nonlinearity; lower temperature measurements to examine the fully gapped regime; and devices constructed to examine systematic variation of transport with crystallographic direction, to probe for anisotropy of field-driven gap breakdown.

\section{Acknowledgements}
W.J.H., H.J, and D.N. gratefully acknowledge support from DOE BES award DE-FG02-06ER46337. Work at UCSB was supported by in part by FAME, one of six centers of STARnet, a Semiconductor Research Corporation program sponsored by MARCO and DARPA, and by a MURI program of the Army Research Office (grant no. W911-NF-09-1-0398).


\end{document}